\documentclass[12pt]{article}
\usepackage{graphicx}

\newsavebox{\sboxpubnumber}
\newsavebox{\sboxpubdate}
\newcommand{\pubdate}[1]{\begin{lrbox}{\sboxpubdate}{#1}\end{lrbox}}
\newcommand{\pubnumber}[1]{\begin{lrbox}{\sboxpubnumber}{\begin{tabular}{l} #1 \\
				 \usebox{\sboxpubdate}
				 \end{tabular}}
                           \end{lrbox}
                           \pubblock}
\newcommand{\Title}[1]{\begin{center} {\Large #1 } \end{center}}
\newcommand{\Author}[1]{\begin{center}{ \sc #1} \end{center}}
\newcommand{\Address}[1]{\begin{center}{ \it #1} \end{center}}

\newcommand{\pubblock}{\rightline{
			\usebox{\sboxpubnumber}}}
\newenvironment{Abstract}{\begin{quotation}  }{\end{quotation}}
\newenvironment{Presented}{\begin{quotation} \begin{center}
             PRESENTED AT\end{center}\bigskip
      \begin{center}\begin{large}}{\end{large}\end{center}
      \end{quotation}}
\newcommand{\Acknowledgements}{\bigskip  \bigskip \begin{center} \begin{large}
             \bf ACKNOWLEDGEMENTS \end{large}\end{center}}

\newcommand{\beq}{\begin{equation}}
\newcommand{\eeq}{\end{equation}}
\newcommand{\beqa}{\begin{eqnarray}}
\newcommand{\eeqa}{\end{eqnarray}}

\newcommand{\del}{\partial}

\newcommand{\deldag}{\mathbin{\partial\mkern-10.5mu\big/}}

\newcommand{\Adag}{\mathbin{A\mkern-10.5mu\big/}}

\newcommand{\nn}{\nonumber}

\def\Slash#1{#1\kern-0.55em\raise.05ex\hbox{/}}
\def\slash#1{#1\kern-0.5em\raise.05ex\hbox{{$\scriptstyle /$}}}

\newcommand{\cpdag}{\mathbin{{\rm CP}\mkern-20.0mu\big/}}
\def\scp{s_{\cpdag}}

\begin{document}

\begin{titlepage}
\pubdate{\today}                                
\pubnumber{NORDITA-2001-91 HE, CERN-TH/2002-09, HD-THEP-02-02} 

\vfill
\Title{Quantum Boltzmann equations for electroweak
       baryogenesis including gauge fields}
\vfill
\Author{Kimmo Kainulainen\footnote{Speaker}$^,$\footnote{University of 
Jyv\"askyl\"a, Finland on leave of absence.}}
\Address{Theory Division, CERN, CH-1211, Geneva 23, Switzerland 
{\rm and} \\ NORDITA, Blegdamsvej 17, DK-2100, Copenhagen, Denmark \O}
\vfill

\vfill
\Author{Tomislav Prokopec,
            Michael G.\ Schmidt, and Steffen Weinstock}
\Address{Institut f\"ur Theoretische Physik, Universit\"at Heidelberg \\
          Philosophenweg 16, D-69120 Heidelberg, Deutschland/Germany}
\vfill

\begin{Abstract}
We review and extend to include the gauge fields our derivation of the 
semiclassical limit of the collisionless quantum transport equations for 
the fermions in presence of a CP-violating bubble wall at a first order 
electroweak phase transition. 
We show how the (gradient correction modified) Lorenz-force appears both 
in the Schwinger-Keldysh approach and in the semiclassical WKB-treatment. 
In the latter approach the inclusion of gauge fields removes the apparent 
phase reparametrization dependence of the intermediate calculations. 
We also discuss setting up the fluid equations for practical calculations
in electroweak baryogenesis including the self-consistent (hyper)electric 
field and the anomaly.
\end{Abstract}
\vfill

\begin{Presented}
    COSMO-01 \\
    Rovaniemi, Finland, \\
    August 29 -- September 4, 2001
\end{Presented}   
\vfill
\end{titlepage}

\def\thefootnote{\fnsymbol{footnote}}
\setcounter{footnote}{0}


\section{Introduction}

Among the models put forward for creation of baryon asymmetry, 
those based on electroweak baryon number 
violation~\cite{KuzminRubakovShaposhnikov} during the electroweak 
phase transition (EWPT) stand out because the physics involved in 
the relevant processes is either being tested, or can soon be tested 
in accelerator experiments.  A good example is the sphaleron 
wash-out constraint: to maintain an asymmetry created in EWPT, 
the baryon number violation must turn off in the broken phase.  
Whether this is the case, depends on the strength of the 
transition~\cite{Shaposhnikov86}, which again depends on the 
model parameters, and can be studied using either effective 
potential or lattice methods. In particular it has been 
shown~\cite{KajantieLaineRummukainenShaposhnikov} that, 
given the current bound on the Higgs mass, the 
transition in the Minimal Standard Model (MSM) does not satisfy
this constraint.  So, to explain the origin of baryons one is 
led to consider extensions of the MSM, and perhaps the most 
natural is to introduce supersymmetry (SUSY). Among SUSY 
extensions of MSM, already the simplest Minimal Supersymmetric 
Standard Model (MSSM)~\cite{CarenaQuirosWagnerETAL} and its next to 
minimal extension 
(NMSSM)~\cite{PietroniDaviesFroggattMoorhouseHuberSchmidt} contain 
additional scalars which can strengthen the phase transition as 
required for baryogenesis.

Because of the out-of-equilibrium requirement~\cite{sakharov}, baryons 
can only be created within the moving phase transition walls or in 
their immediate vicinity during EWPT.  It is no surprise then that the 
ratio of the wall width $\ell_{\rm wall}$ to the other characteristic 
scales, such as the scattering lengths $\ell_{\rm scatt} \sim 1/\Gamma$ 
and de Broglie wave length $\ell_{\rm dB} = 1/p \sim 1/T$ of the 
particles involved, strongly influences the mechanism of baryogenesis. 
The wall width also controls the CP-violation, which in general is 
introduced through complex phases in the mass matrices. When mass 
matrix elements depend on the higgs condensate, the phases of mass
eigenstates can change over the wall profile influencing the propagation 
of particles and antiparticles in different ways. If the wall is wide in 
comparison with the typical de Broglie wave length $\ell_{\rm wall} 
\ll \ell_{\rm dB}$, these effects should be computable using 
semiclassical (SC) methods~\cite{JPT-thick}.

In supersymmetric models the bubble walls are found to be quite slow 
and thick~\cite{MooreProkopec95,MorenoQuirosSeco}, so that the 
semiclassical approach~\cite{JPT-thick} 
should be applicable. It has been used to compute the baryogenesis from 
chargino sector of MSSM~\cite{ClineJoyceKainulainen,
ClineJoyceKainulainen-II,HuberJohnSchmidt-I} 
and in the NMSSM~\cite{HuberSchmidt}. However, the semiclassical 
approach is based on the {\em assumption} that the plasma in a spatially 
varying background can be described by a collection of single particle 
WKB-excitations.  Despite the obvious intuitive appeal, this assumption 
was lacking a fundamental proof until recently~\cite{kpsw-I}.  
Proving the validity of the SC-approach was important, in particular 
because many other formulations of the transport equations in the limit 
of slowly varying walls had been but forward, which gave both 
quantitatively and qualitatively different results from the SC 
predictions~\cite{OtherApproaches}. Moreover, there was some initial 
confusion in identifying the correct physical excitations and their 
distribution functions even within the SC 
picture~\cite{JPT-thick,ClineJoyceKainulainen}, and a 
fully consistent SC-computation was introduced only in 
ref.~\cite{ClineJoyceKainulainen-II}.

Here we review our derivation of collisionless SC-transport equations 
for fermions with spatially varying CP-violating mass terms~\cite{kpsw-I}
starting from Schwinger-Keldysh equations for the dynamical two point 
Green function (Wightman function) $G^<(u,v)$. Moreover, the analysis 
of~\cite{kpsw-I} is extended to include the gauge fields, which give 
rise to the usual Lorenz-force on gauge-charged particles. We also 
generalize the WKB-analysis of ref.~\cite{ClineJoyceKainulainen-II} 
to include the gauge fields, and show how the gauge degree of freedom is 
connected to the ``reparametrization dependence'' of canonical variables 
in the WKB-picture. As in~\cite{kpsw-I} we concentrate on the 
1+1-dimensional model; the full 3+1 dimensional case will be considered 
in ref.~\cite{KPSW2}.

The most remarkable result of our analysis is that to first order in 
gradients (or equivalently in $\hbar$), the equations of motion admit
a {\em spectral decomposition solution} for $G^<$ in terms of on-shell 
quasiparticle excitations; this qualitatively proves the validity
of the above mentioned assumption of single particle picture underlying 
the SC-approach.  Moreover, the on-shell momenta set by a dispersion 
relation derived from the equations of motion quantitatively agree 
with the SC results of~\cite{ClineJoyceKainulainen-II} (and also with
an earlier Schwinger-Dyson equations based 
calculation~\cite{JKP-FermionProp}). 
Finally, the on-shell distribution functions are defined for the 
quasiparticle excitations, and are shown to obey the usual kinetic 
(Vlasov) equations, with the SC group velocity $v_{s\pm} \equiv 
k_z/\omega_{s\pm}$ and the force 
$F_{s\pm} = \omega_{s\pm} dv_{s\pm}/dt$.  

However, the equations of motion only admit a spectral decomposition 
solution to first order in $\hbar$.  At higher orders more complicated 
structures arise and the SC-picture of a plasma consisting of single 
particle excitations breaks down. Quite fortuitiously for the 
EWBG-program this is all we need because, as we shall see, the 
dominant CP-violation terms appear precisely at order $\hbar$.

This contribution is organized as follows. In section 2 we derive the 
SC WKB-equations for a single Dirac fermion with a spatially varying
complex mass term including the gauge field. In section 3 we derive
the SC kinetic equation from Schwinger-Keldysh formalism, and in 
section 5 we study vector and axial vector divergences, and show how 
one sets up moment equations for practical baryogenesis computations. 
Finally, section 5 contains a discussion and summary.

\section{WKB method}
\label{WKB method}

We will here derive the semiclassical dispersion relation for a fermionic 
field with a complex spatially varying mass term. More precisely, we take 
our system to be described by the effective lagrangian
\beq
   {\cal L} = i \bar{\psi}(\deldag + ie\Adag )\psi - \bar{\psi}_Lm\psi_R
              - \bar{\psi}_Rm^*\psi_L + {\cal L}_{\rm int}
\,,
\label{lagrangian}
\eeq
where $A_\mu$ is the $U(1)$-gauge field, ${\cal L}_{\rm int}$ contains 
(here unspecified) interactions and
\beq
      m(z) = m_R(z) + i m_I(z) = |m(z)|\mbox{e}^{i\theta(z)}
\label{mass1}
\eeq
is a mass term arising from an interaction with some CP-violating scalar 
field condensate with a planar symmetry. Planar symmetry effectively 
reduces the problem to 1+1 dimensions\footnote{
In partiuclar the gauge field components $A_{x,y}$ can be taken to
vanish due to planar symmetry.} 
(by use of boosts), and to keep 
the notation simple and consistent with the rest of the paper, we will 
assume this to be strictly the case here.

The basic assumption in the 
semiclassical treatment is that the plasma can be approximated by a 
collection of single particle excitations, whose dispersive properties 
can be studied by employing the WKB-methods. 
The lagrangian~(\ref{lagrangian}) implies the following Dirac equation 
for the spinor wave function $\psi(z,t)$:
\beq
\left(i{\deldag} - e \Adag - m_R - i\gamma^5 m_I \right) \psi = 0.
\label{wkb.0}
\eeq
We are interested in stationary solutions of the form  $\psi (t,z) = 
e^{-ik_0t}\psi(k_0;z)$.  Eq.~(\ref{wkb.0}) then becomes
\beq
\left( \gamma^0 \kappa_0  +  i \gamma^3 (\del_z - ieA_z) 
     - m_R - i\gamma^5 m_I \right) \psi(k_0; z) = 0,
\label{wkb.1}
\eeq
where $\kappa_0 \equiv k_0 - eA_0$. Here spin is also a good quantum 
number and we can write the wave function as a direct product of chiral
and spin degrees of freedom:
\beq
  \psi \equiv  
  \left(\begin{array}{c} L_s \cr R_s \end{array} \right) \otimes \chi^s,
\label{wkb.3}
\eeq
where $\chi^s$ are the eigenspinors of the spin operator $\sigma^3$. 
Eq.~(\ref{wkb.1}) then splits to two coupled equations or the left and 
right chiral wave functions:
\beqa
  (\kappa_0 - e s A_z - i s\partial_z) L_s &=& m R_s
\nonumber\\
  (\kappa_0 + e s A_z + i s\partial_z) R_s &=& m^* L_s,
\label{wkb.4}
\eeqa
One can use, say, the first of equations (\ref{wkb.4}) to eliminate $R_s$ 
from the second one, to obtain a second order equation for $L_s$. That 
equation can be analyzed by making the WKB-parametrization for $L_s$: 
\beq 
   L_s \equiv u_s e^{i\int^z k_s dz'}.
\label{wkb.7a}
\eeq
In this way one finds the following two real valued equations
\beqa
 \kappa_0^2 - \kappa_3^2 - |m|^2  + (s\kappa_0 + \kappa_3)\theta' 
  + \frac{u_s''}{u_s} - \frac{|m|'}{|m|} \frac{u_s'}{u_s}  &=& 0
\label{wkb.8}
\\
 \kappa_3' + (2\kappa_3 - \theta') \frac{u_s'}{u_s} 
   - (s\kappa_0 + \kappa_3)\frac{|m|'}{|m|} &=& 0,
\label{wkb.9}
\eeqa
where $\kappa_3 \equiv k_s - eA_z$.  These equations are exactly 
analogous to the ones obtained in~\cite{ClineJoyceKainulainen-II}, 
with the sole exception that the bare momentum 
of~\cite{ClineJoyceKainulainen-II}, has here been replaced by the 
one including the gauge field:
\beq 
  k_\mu \rightarrow \kappa_\mu  \equiv  k_\mu - e A_\mu .
\eeq
In absence of the gauge field, the WKB-equations 
of~\cite{ClineJoyceKainulainen-II} were variant under phase 
redefinitions of the $\psi$-field, which led to the ``reparametrization
variance" of the canonical momentum.  Equations (\ref{wkb.8}) and 
(\ref{wkb.9}) are clearly invariant under complete gauge transformations 
$\psi \rightarrow e^{i\alpha}\psi$ and $A_\mu \rightarrow A_\mu - 
(i/e)\partial_\mu \alpha$, and the reparametrization variance 
of~\cite{ClineJoyceKainulainen-II} is here seen to correspond to 
the usual freedom of adding any four divergence of a scalar 
function $\partial_\mu \phi$ to the gauge potential $A_\mu$.

While complex in appearance, equations (\ref{wkb.8}-\ref{wkb.9}) 
have the advantage of being readily solved iteratively in gradient 
expansion. This part of the analysis is analogous to one shown 
in~\cite{ClineJoyceKainulainen-II}, and hence we only quote the final
results. To the first order in gradients the dispersion relation
reads
\beq
  \kappa_3 \; \equiv \;  k_s - eA_\mu \;
    =  \; p_0   +  \scp \frac{(s k_0 + p_0)\theta'}{2p_0} ,
\label{wkb.10}
\eeq
where $p_0 = {\rm sign}(p_s)\sqrt{k_0^2 - |m|^2}$ and 
$\scp = 1 (-1)$ for particles (antiparticles). Equation 
(\ref{wkb.10}) can be inverted to give the energy in terms of
the momentum variable $\kappa_3$:
\beq
\kappa_0 \simeq 
       \sqrt{(\kappa_3 - \scp \frac{s\theta'}{2})^2 + |m|^2}
           - \scp \frac{s\theta'}{2}.
\label{wkb.11}
\eeq
Now it is easy to compute the physical momentum, which is defined as 
usual, in terms of the group velocity 
$k_z \equiv \kappa_0 v_g = \kappa_0 (\partial_{\kappa_3}\kappa_0)_z$. 
After some algebra one finds
\beq
  k_z = p_0   +  \scp \frac{s|m|^2\theta'}{2p_0\kappa_0}.
\label{wkb.12}
\eeq
Finally, the physical force is defined as $F_z = \dot k_z$. The 
only difference between the computation here and in 
ref.~\cite{ClineJoyceKainulainen-II} is that in the presence of an 
electric field the energy $\kappa_0$ is not conserved; indeed after 
some algebra one finds the expected result 
$\dot \kappa_0 = e v_g E_z$, where $E_z$ is the electric field 
in the $z$-direction. Similarly, the classical force is found to be
\beq
F_z  \; = \; \dot \kappa_0 v_g + \kappa_0 \dot v_g 
     \; = \;  - \frac{{|m|^2}^{\,\prime} }{2\kappa_0}
     + \scp \frac{s(|m|^2\theta^{\,\prime})^{\,\prime}}{2\kappa_0^2}
     + e E_z \left( 1 - \scp \frac{s|m|^2\theta'}{2\kappa_0^3} \right).
\label{voima}
\eeq
We have here computed all expressions to the first order in gauge field 
and to the second order in gradients.  If the electric field $E_z$ is a 
self-consistent field created by the charge separation due to 
CP-violating force in (\ref{voima}), it is formally a second order term
and the gradient correction multiplying $eE_z$ should be dropped.
According to the semiclassical assumption the Vlasov equation for the 
phase space density of collisionless WKB-excitations can now be written 
as
\beq
  \partial_t f_s + v_g \partial_z f_s  + F_z \partial_{k_z} f_s = 0.
\eeq
The above computation is improvement upon 
ref.~\cite{ClineJoyceKainulainen-II} in that we include also the 
self consistent electric field which is necessarily generated by the
charge separation induced by the wall. It also better explains the
phase reparametrization variance observed in connection with the 
canoncial momentum.  The WKB-derivation however depends on the
important assumption that the plasma in a spatially varying 
background can be described in a single particle WKB-picture. To
determine whether this really is so, and to which accuracy, one has
to go beyond the semiclassical approximation. This is what we do in 
the next section.

\section{Schwinger-Keldysh method}

The statistical properties of an out-of-equilibrium system can be
described by the two point Wightman function~\cite{sdoldies}
\beq
  G^<_{\alpha\beta}(u,v) 
   = i \langle \bar\psi_\alpha(u) 
                e^{ie\int^u_v dy \cdot A}
               \psi_\beta(v) 
       \rangle ,
\label{G-less}
\eeq
where $\left<\cdot \right>$ denotes the expectation value with respect 
to the initial state and the path for the gauge flux is defined as
\beq
  \int^u_v dy \cdot A  \equiv \int^{1/2}_{-1/2} ds \, r^\mu A_\mu(x+sr),
\label{gflux}
\eeq
where $r\equiv u-v$ denotes the relative and $x=(u+v)/2$ the 
center-of-mass coordinate. The definition (\ref{G-less}) is a 
gauge-invariant generalization~\cite{HeinzZhuang} of the 2-point 
function used in~\cite{kpsw-I}. $G^<$ corresponds to the off-diagonal 
part of the fermionic two-point function in the Schwinger-Keldysh 
formalism~\cite{SchwingerKeldysh}, which indeed is the method of choice 
to derive the equations of motion for $G^<$ including interactions. 
However, for noninteracting system studied here, all one needs is the 
familiar Dirac equation. Using Eq.~(\ref{wkb.0}) one finds:
\beq
\left( i \Slash D_u - m(u)^* P_L - m(u) P_R 
      + e \gamma^\mu \frac{\partial}{\partial u_\mu} \int_v^u dx \cdot A 
\right) G^<(u,v) = 0.
\label{DirectRep}
\eeq
In order to study equation~(\ref{DirectRep}) in gradient expansion we 
performa a Wigner transform of $G^<$ to the mixed representation:
\beq
G^<(x,k) \equiv \int d^{\,4} r \, e^{ik\cdot r} G^<(x + r/2,x-r/2),
\label{wigner1}
\eeq
where $r$ and $x$ are defined as in (\ref{gflux}). In the mixed 
representation the equation of motion becomes
\beq
 \left( \Slash \Pi(k,x) + \frac{i}{2} \Slash D(k,x)   
        -  m_R(x-\frac{i}{2}\partial_k)
        -i m_I(x-\frac{i}{2}\partial_k)\gamma^5  
 \right) G^<(k,x) = 0.
\label{WignerRep21}
\eeq
where
\beq
  m_{R,I} (x-\frac{i}{2}\partial_k) \equiv  m_{R,I} (x)
  e^{-\frac{i}{2}\stackrel{\!\!\leftarrow}{\partial_x}\cdot\,\partial_k}
\label{massexp}
\eeq
and the gauge generalized derivative and momentum terms are (in 
agreement with Zhuang and Heinz~\cite{HeinzZhuang}):
\beqa
  D_\mu(x,k)   &\equiv& \partial_\mu 
   - e\int^{+1/2}_{-1/2} ds F_{\mu\nu}(x-is\partial_k) 
                             \partial_{k_\nu} \nn \\ 
  \Pi_\mu(x,k) &\equiv& k_\mu-ie\int^{+1/2}_{-1/2}ds 
                          s F_{\mu\nu}(x-is\partial_k) 
                              \partial_{k_\nu}.
\label{DPi}
\eeqa
Above equations of motion are manifestly gauge invariant, and one 
can in fact always rewrite the $F_{\mu\nu}$-terms in any of the above 
formulas in terms of physical electric and magnetic fields. Using 
$F_{0i} = E_i$ and $F_{ij} = -\epsilon_{ijk} B_k$ one finds
\beq
\gamma^\mu F_{\mu\nu} \partial_{k_\nu} = 
        - \gamma^0 \vec E \cdot \partial_{\vec k} 
        + \vec \gamma \cdot \vec E \, \partial_{k_0}
        + \vec \gamma \cdot ( \vec B \times \partial_{\vec k} ).
\label{EBfieldExp}
\eeq

The crucial advantage of the representation (\ref{wigner1}) is that it 
separates the internal fluctuation scales, described by momenta $k$, from 
the external ones which show up as a dependence of $G^<$ on the center-of-mass
coordinate $x$, and thus gives us the chance of exploiting possible hierarchies 
between these scales.  Indeed, when the external fields vary slowly, one can
truncate the Taylor series operator expansions of the mass and gauge field 
operators appearing in (\ref{massexp}-\ref{EBfieldExp}) to some finite 
order in gradients (or equivalently in powers of $\hbar$). To the lowest
order we get
\beqa
  D_\mu(x,k)   &\simeq& \partial_\mu  
                      - e \, F_{\mu\nu}(x) \partial_{k_\nu} \nn \\ 
  \Pi_\mu(x,k) &\simeq& k_\mu.
\label{lowest2}
\eeqa
In the planar symmetric case of an electroweak bubble wall, there can be no 
magnetic field in the plasma frame. Also $G^<$ can only depend on the spatial 
coordinate orthogonal to the wall, $z\equiv x^3$.  Continuing to work in the 
strictly 1+1-dimensional case like in the previous section, we also drop the 
$k_{||}$-term. With these simplifications the equation (\ref{WignerRep21}) 
reduces to
\beq
\left({\hat k}_0 + {\hat k}_z \gamma^0 \gamma^3
          - {\hat m}_0\gamma^0 + i {\hat m}_5 \gamma^0\gamma^5 \right)
           i \gamma^0G^< = 0 ,
\label{G-less-eom2}
\eeq
where we used the following shorthand notations
\beqa
{\hat k}_0 &\equiv& k_0 + \frac{i}{2} (\partial_t + eE_z \partial_{k_z})
\nn \\
{\hat k}_z &\equiv& k_z - \frac{i}{2} (\partial_z + eE_z \partial_{k_0})
\label{shorts}
\eeqa
where $E_z$ is the electric field in the $z$-direction, and
\beq
{\hat m}_{0(5)} \simeq m_{R(I)}(z) - \frac{i}{2} m_{R(I)}' \partial_{k_z}.
\label{shorts2}
\eeq
Apart from the electric field $E_z$ appearing in the definitions 
(\ref{shorts}), equation (\ref{G-less-eom2}) is identical to the one we 
studied in~\cite{kpsw-I}. Hence the further analysis proceeds in very 
a similar fashion, we can skip many of the details here. Due to 
conservation of spin in $z$-direction $G^<$ can be decomposed as 
follows:
\beq
  -i\gamma^0G^<_s = \frac12(1+s\sigma^3)\otimes g^<_s,
\label{connection}
\eeq
where $\sigma^3$ is the usual Pauli matrix. Using this structure, and 
making the appropriate identifications $\gamma^0 \rightarrow \rho^1$, 
$-i\gamma^0\gamma^5  \rightarrow \rho^2$ and $-\gamma^5 \rightarrow 
\rho^3$, the original $4\times4$ problem reduces to a two-dimensional 
one:
\beq
\left( {\hat k}_0 - s {\hat k}_z\rho^3
        - \rho^1 {\hat m}_0 - \rho^2 {\hat m}_5  \right) g^<_s = 0,
\label{Gs-eom}
\eeq
where $\rho^i$ are the Pauli matrices, which span the remaining chiral
indices. The hermitean matrices $g^<_s$ can be expanded as
\beq
g^<_s \equiv \frac12 \left( g^{s}_0 +  g^{s}_i \rho^i \right) ,
\label{glesss}
\eeq
where the signs and normalizations are chosen such that $g_0^{s}$ 
measures the number density of particles with spin $s$ in phase space. 

Equation~(\ref{Gs-eom}) still constitutes a set of four coupled complex 
differential equations for components of $g^{s}_a$ ($a=0,i$). They can be
obtaned by multiplying (\ref{Gs-eom}) successively by $1$ and $\rho^i$ 
and taking the trace:
\def\hi{\phantom{i}}
\beqa
{\hat k}_0 g^{s}_0  - \hi s{\hat k}_z g^{s}_3
            - \hi {\hat m}_0 g^{s}_1 - \hi {\hat m}_5 g^{s}_2 &=& 0
\label{eq-sigma1} \\
{\hat k}_0 g^{s}_3  - \hi s{\hat k}_z g^{s}_0
            - i   {\hat m}_0 g^{s}_2 + i   {\hat m}_5 g^{s}_1 &=& 0
\label{eq-sigma2} \\
{\hat k}_0 g^{s}_1  + i   s{\hat k}_z g^{s}_2
            - \hi {\hat m}_0 g^{s}_0 - i   {\hat m}_5 g^{s}_3 &=& 0
\label{eq-sigma3} \\
{\hat k}_0 g^{s}_2  - i   s{\hat k}_z g^{s}_1
            + i   {\hat m}_0 g^{s}_3 - \hi {\hat m}_5 g^{s}_0 &=& 0.
\label{eq-sigma4}
\eeqa
The real parts of these equations contain no time derivatives and hence 
are called {\em constraint equations} (CE). Their very important role is 
to constrain the available space of solutions for the four dynamical, 
{\em kinetic equations} (KE), which correspond to the imaginary parts 
of~(\ref{eq-sigma1}-\ref{eq-sigma4}).
%

\subsection{Constraint equations}

We first use three out of the four constraint equations to iteratively 
solve $g^s_i$ in terms of $g^s_0$ and its derivatives. These 
expressions are then used to eliminate $g^s_i$ from the last remaining 
CE. The most remarkable property of this equation is that all terms 
proportional to $\partial_{k_z} g^{s}_0$, also those involving the 
field $E_z$, cancel to first order in gradients (or $\hbar$). As a 
result $g^{s}_0$ satisfies an {\em algebraic} equation
\beq
\left( k_0^2 -  k_z^2 - |m|^2  + \frac{s}{k_0}|m|^2\theta^{\,\prime}
       \right) g^{s}_0 = 0,
\label{constraint-g0}
\eeq
which admits the spectral solution 
\beqa
g^{s}_0  &=& \sum_\pm \frac{\pi }{2 Z_{s\pm}}
                 \, n_s \,\delta(k_0 \mp \omega_{s\pm})\,.
\label{spectral-dec}
\eeqa
Here $n_s(k_0,k_z,z)$ are some nonsingular functions, the on-shell 
energies $\omega_{s\pm}$ are specified by the roots of equation
\beq
\Omega^2_s \; \equiv \;
   k_0^2 - k_z^2 - |m|^2 + \frac{s}{k_0}|m|^2\theta^{\,\prime} 
    \; = \; 0,
\label{DR1}
\eeq
and the normalization factor is defined as $Z_{s\pm} \equiv 
\frac{1}{2 \omega_{s\pm}\,} |\partial_{k_0}
\Omega^2_s|_{k_0=\pm\omega_{s\pm}}$.

The solution (\ref{spectral-dec}) is in many ways the most important 
result of our analysis, because it precisely proves the WKB-assumption 
that the plasma, to a certain accuracy, can be described by a collection 
of single particle on-shell excitations.  Below we shall further show 
that also the WKB-dispersion relation and the semiclassical equations 
of motion of ref.~\cite{ClineJoyceKainulainen-II} do arise in this 
picture. It is important to note however, that the confinement to sharp 
energy shells is only valid up to first order in gradients; beyond 
first order the constraint equation for $g_0^s$ is no longer algebraic 
and hence does not admit a spectral solution, implying breakdown of
the SC picture.

We end this subsection noting that, when treated to the first order in 
gradients, Eqn.~(\ref{DR1}) yields the solution
\beq
     \omega_{s\pm} = \omega_0 \mp s \frac{|m|^2\theta'}{2\omega^2_0}\,,
     \qquad \qquad \omega_0 = \sqrt{ k_z^2 + |m|^2}.
\label{dispersion1}
\eeq
Inverting this to give $k_z$ in terms of energy, one finds it to be 
identical with the WKB-relation (\ref{wkb.12}) for the physical 
momentum. 

\subsection{Kinetic equations}

We now turn our attention to the kinetic equations. One could 
choose any one out of the four equations to describe the dynamics 
along with the constraint equations. The quantity of most direct 
interest is $g^{s}_0$ however, since it carries information on the 
particle density in phase space.  From (\ref{eq-sigma1}) one gets
\beq
      (\partial_t + eE_z\partial_{k_z}) g^{s}_0 
   + s(\partial_z + eE_z\partial_{k_0}) g^{s}_3
            - m_R' \partial_{k_z}g^{s}_1
            - m_I' \partial_{k_z}g^{s}_2 = 0.
\label{KEA}
\eeq
Using constraint equations to eliminate $g^{s}_i$ as in the 
previous section one finds
\beqa
&&\del_t  g^s_0 + \frac{k_z}{k_0}\partial_z g^s_0
         + eE_z \partial_{k_0} \left(\frac{k_z}{k_0} g^s_0 \right) 
\nn \\ && \hskip 0.80 cm 
       - \left(  \frac{|m|^{2{\,\prime}}}{2k_0}
               - \frac{s}{2k_0^2}(|m|^2\theta')^{\,\prime}
               - eE_z \left( 1-\frac{s|m|^2\theta'}{2k_0^3} \right) 
         \right) \partial_{k_z}g^s_0 = 0.
\label{boltzmann1}
\eeqa
We have so far used three out of four constraint equations. 
The fourth constraint (\ref{spectral-dec}) forces the solutions to 
the single particle on-shell, and is easily accounted for by inserting 
the spectral solution (\ref{spectral-dec}) into (\ref{boltzmann1}) and 
then separately integrating over the positive and negative frequencies 
$k_0$.  The total divergence term with respect to $k_0$ drops and get 
the familiar form for the Vlasov equation
\beq
    \del_t f_{s\pm}
      + v_{s\pm} \del_z f_{s\pm}
      + F_{s\pm} \partial_{k_z} f_{s\pm} = 0,
\label{boltzmann1-sc}
\eeq
where $f_{s+} \equiv n_s(\omega_{s+},k_z,z)$ and 
$f_{s-} \equiv 1 - n_s(-\omega_{s-},-k_z,z)$ are the particle and
antiparticle distribution functions with spin $s$, respectively, 
the quasiparticle group velocity is given by $v_{s\pm} \equiv 
k_z/\omega_{s\pm}$ and the spin-dependent and CP-violating 
{\em semiclassical force} reads
\beq
F_{s\pm} = -\frac{{|m|^2}'}{2\omega_{s\pm}}
             \pm \frac{s(|m|^2\theta')'}{2\omega_{s\pm}^2} 
       + eE_z \left( 1 \mp \frac{s|m|^2\theta'}{2\omega_{s\pm}^3} 
              \right) .
\label{scforce1}
\eeq
in complete analogy with the semiclassical result. The physcial 
significance of (\ref{scforce1}) is that due to derivative (quantum)
corrections the spin degeneracy is lifted at first order in gradients 
and hence particles (antiparticles) of different spin experience 
different accelerations in a spatially varying background.

\section{Continuity equations and sources for BG}
\label{Cont-sources}

We now investigate how the CP-violating sources appear in the equations 
for the divergences of the vector and axial vector currents. First note
that any current operator can be computed as a trace in the Wigner 
representation, weighted by $G^<$:
\beq
 \langle \bar\psi(x) X^\mu \psi(x) \rangle 
   = \int \frac{d^{\,2}k}{(2\pi)^2} {\rm Tr}(-iG^< X^\mu).
\label{vecdiv0}
\eeq
For vector current $X^\mu = \gamma^\mu$ and for axial one $X^\mu = 
\gamma^\mu \gamma^5$. Using the above expressions and constraint 
equations from (\ref{eq-sigma1}-\ref{eq-sigma4}) one can verify that
the divergence of the vector current is conserved and the axial one
is broken by the complex mass as expected:
\footnote{Note that the sign of the $m_R$-term in the axial current
divergence was incorrect in our original work~\cite{kpsw-I}. That
error was compensated however, by a another trivial sign error in 
the definition of the pseudoscalar density in terms of $g^s_2$.}
\beqa
\partial_\mu j^\mu &=& 0 
\label{vecdiv1}
        \\
\partial_\mu j^\mu_{5} &=& 
           2iM_R \langle\bar{\psi}\gamma^5\psi \rangle
           - 2M_I \langle\bar{\psi} \psi \rangle,
\label{axdiv1}
\eeqa
where $\langle\bar{\psi}_s\psi_s\rangle = \int_k g^s_1$ denotes the 
scalar, and $\langle\bar{\psi}_s\gamma^5\psi_s\rangle = - i \int_k g^s_2$ 
the pseudoscalar density, and $\int_k \equiv \int d^2k/(2\pi)^2$.
If we define the fluid density\footnote{
The fluid density $n_{s\pm}$ should not be confused with the phase 
space density $n_s$ in~(\ref{spectral-dec}).}
$n_{s\pm}$ and fluid velocity moments as\footnote{
Here we defined integration over the positive frequencies. To obtain 
equations for antiparticles one integrate over negative frequencies
taking into account the appropriate thermal equilibrium limit as in
derivation of (\ref{scforce1}).}
\begin{eqnarray}
  n_{s+}  &\equiv& \int_+\frac{d^{\,2}k}{(2\pi)^2} g^{s}_0
\nonumber\\
  \langle v_{s+}^p \rangle &\equiv& 
        \frac{1}{n_{s+}} \int_{+}\frac{d^{\,2}k}{(2\pi)^2}
        \left( \frac{k_z}{k_0} \right)^p g^{s}_0 \,,
\label{Ncont-2}
\end{eqnarray}
then Eqs.~(\ref{vecdiv1}-\ref{axdiv1}) can be written as:
\beqa
   \del_t n_{s\pm}
    + \partial_z(n_{s\pm} \langle v_{s\pm}\rangle) &=& 0,  
\label{Ncont1}   
\\
   \partial_t \left(n_{s\pm} \langle v_{s\pm} \rangle  \right)
     + \del_z \left(n_{s\pm} \langle v_{s\pm}^2 \rangle \right)
     &=&  eE_z |m|^2 {\cal I}_{3s+} + S_{s\pm},
\label{Ncont2}
\eeqa
where the quantity $S_{s\pm}$ is related to the baryogenesis 
source\footnote{Note that the equations 
(\ref{Ncont1}-\ref{Ncont2}) are somewhat formal, and require more 
work before useful for practical computations. In particular the 
velocity-dependence of the actual source only comes about after one 
expands various quantities in the wall velocity around the equilibrium 
distribution. For details see~\cite{ClineJoyceKainulainen-II} 
and~\cite{KPSW2}.}
\cite{ClineJoyceKainulainen-II}, 
and is given by the average over the semiclassical force 
(\ref{scforce1}) divided by $k_0$:
\beq
    S_{s\pm} =
    - \frac 12 \vert m \vert^{2\,\prime} {\cal I}_{2s\pm}
    + \frac 12 \scp \, s(\vert m \vert^{2}\theta')' \, {\cal I}_{3s\pm}.
\label{realsource}
\eeq
with
\beq
{\cal I}_{ps+} \equiv 
 \int_{+}\frac{d^2 k}{(2\pi)^2} \frac{g_{0i}^s}{k_0^p}.
\label{iota}
\eeq
In anticipation of the fact that $E_z$ is a self-consistent field, 
we dropped the gradient corrections to the Lorenz force above. The 
$|m|^2$-suppression in the $E_z$-term is particular to 1+1 dimensions,
and to our use of velocity moments in fluid equations as in 
ref.~\cite{ClineJoyceKainulainen-II}. In the actual baryogenesis
computation one expects that the role of the electric field is 
expected to be taken by the hyperelectric field corresponding to 
the hypercharge $U(1)_B$ symmetry.

From expressions (\ref{Ncont1}) and (\ref{Ncont2}) it is obvious that
the vector current divergence corresponds to the zeroth and the axial
current divergence to the first velocity momentum of the kinetic
equation (\ref{scforce1}). The former provides just the usual continuity 
equation, and the nontrivial CP-violating source and the  electric field 
only appear in the latter. 
This connection with the axial current nicely explains why the 
semiclassical source appeares only at first order in moment expansion 
in earlier treatments~\cite{JPT-thick,ClineJoyceKainulainen-II}. 

Equations (\ref{Ncont1}-\ref{Ncont2}) do not close, because the field 
$E_z$ is not yet determined and since the two equations contain three 
other kinetic unknowns $n_{s\pm}$, $\langle v_{s\pm} \rangle$ and 
$\langle v^2_{s\pm} \rangle$. The former problem can be 
handled~\cite{ClineKainulainen} by adding the Gauss law
\beq
\partial_z E_z = e \sum_s (n_{s+} - n_{s-}),
\label{gausslaw}
\eeq
which completely defines the self-consistent electric field. The 
latter problem is a standard one for the moment expansions and requires 
a truncation approximation. For example one may set 
$\langle v^2_{s\pm} \rangle \rightarrow \langle v_{s\pm} \rangle^2 
+ \sigma^2_{s\pm}$ and use the equilibrium distributions for 
$f_{s\pm}$'s to evaluating the variance $\sigma^2_{s\pm} \equiv 
\langle v_{s\pm}^2 \rangle - \langle v_{s\pm}\rangle^2$ and the 
integrals ${\cal I}_{ps\pm}$  appearing in the source term 
(\ref{realsource}).

The first source-term in (\ref{realsource}) is primarily relevant for 
phase transition dynamics~\cite{MooreProkopec95,MorenoQuirosSeco},
but it also contains a spin dependent and CP-violating part coming 
from the wave function renormalization~\cite{kpsw-I}. That term, 
together with the second source provide the charge separation necessary 
for baryogenesis. To promote equations ~(\ref{Ncont1}-\ref{Ncont2}) 
and~(\ref{gausslaw}) into  transport equations for baryogenesis 
calculations we still need to generalize them to include collisions. 
For a discussion of this problem see the talk of Tomislav Prokopec 
in these proceedings~\cite{tomproc}. 

Let us finally discuss the role of the gauge anomaly in the fluid
equations. In the 1+1-dimensional case the U(1)-theory is anomalous, 
and in general the anomaly~\cite{peskinsch}
\beq
  (\partial_\mu j^\mu_5)_{\rm anomalous} =
   - \frac{e}{2\pi} \epsilon_{\mu\nu}F^{\mu\nu}
\eeq
should be added to the axial vector divergence (\ref{axdiv1}).
Because the axial current divergence, apart from the anomaly, has 
been here shown to coincide with the first velocity moment of the
fluid equations~\cite{kpsw-I}, we can draw the following corollary: 
the first two moments of the fermionic transport equations, with the 
anomalous term added to the latter, together with the appropriate 
dynamical Langevin equations of motion for the gauge fields, provides 
a complete and self-consistent method for treating the CP-, P- and
fermion number violating (baryon number violating) out-of-equilibrium
interactions required for electroweak baryogenesis.

\section{Discussion and summary}

The question of a first principle derivation of CP-violating fluxes 
in transport equations has been the main theoretical challenge of 
recent work on electroweak baryogenesis. In this talk we reviewed our 
derivation~\cite{kpsw-I} of the kinetic equations appropriate for 
EWBG in a systematic and controlled gradient expansion from 
Schwinger-Keldysh equations for two point Wightmann function, 
and extended it to include the gauge fields.  We also generalized
the derivation of WKB-dispersion 
relations~\cite{ClineJoyceKainulainen-II} to include the gauge
degree of freedom.

We have shown that to first order in $\hbar$ the collisionless 
kinetic equations for fermions allow a single particle description 
with a distribution function which obeys a familiar Vlasov equation
where the group  velocity and the semiclassical force terms contain 
all relevant quantum information, and in particular the CP-violating 
terms which source baryogenesis. 
These results agree with the SC equations derived in the WKB-picture  
in Ref.~\cite{ClineJoyceKainulainen-II},  originally developed 
for EWBG problem in~\cite{JPT-thick}.  A novel new 
feature discovered in our approach is that the semiclassical 
limit is {\em only} valid to first order in $\hbar$. Fortunately
this is not a serious drawback to the EWBG-program, as it is 
just the first order term that carries the dominant source for
baryogenesis.

For simplicity we have considered only the collisionless limit in 1+1 
dimensions. One can show~\cite{KPSW2} that generalization to the 3+1
dimensional case does not affect results qualitatively, and some 
aspects of collision terms are considered by Tomislav Prokopec in 
these proceedings. For the lack of space within this presentation 
we also restricted us to the case of a single Dirac fermion. For the 
case of mixing fermionic fields see ref.~\cite{kpsw-I}.

\Acknowledgements

KK wishes to thank Mikko Laine and Kari Rummukainen for valuable
discussions.

\end{document}